\documentclass[12pt,preprint]{aastex}
\usepackage{lscape}

\shorttitle{Host Galaxies of Long GRBs}
\shortauthors{Levesque}
\slugcomment{\today}

\begin{document}
\title{The Host Galaxies of Long-Duration Gamma-Ray Bursts}
\author{Emily M. Levesque$^1$}
\affil{CASA, Department of Astrophysical and Planetary Sciences, University of Colorado 389-UCB, Boulder, CO 80309, USA}

\begin{abstract}
Long-duration gamma-ray bursts (LGRBs) are the signatures of extraordinarily high-energy events occurring in our universe. Since their discovery, we have determined that these events are produced during the core-collapse deaths of rare young massive stars. The host galaxies of LGRBs are an excellent means of probing the environments and populations that produce their unusual progenitors. In addition, these same young stellar progenitors makes LGRBs and their host galaxies valuable potentially powerful tracers of star formation and metallicity at high redshifts. However, properly utilizing LGRBs as probes of the early universe requires a thorough understanding of their formation and the host environments that they sample. This review looks back at some of the recent work on LGRB host galaxies that has advanced our understanding of these events and their cosmological applications, and considers the many new questions that we are poised to pursue in the coming years.
\end{abstract}
\footnotetext[1]{Hubble Fellow; {\tt Emily.Levesque@colorado.edu}}

\section{Introduction}
Gamma-ray bursts were serendipitously discovered in the late 1960s by the {\it Vela} nuclear test detection satellites (Klebesadel et al.\ 1973) . Subsequent satellite observatories, including the Burst and Transient Source Explorer (BATSE) of the Compton Gamma Ray Observatory (Fishman et al.\ 1989) and the BeppoSAX X-ray astronomy satellite (Boella et al.\ 1997), combined with the work of ground-based observer follow-up of satellite detections, determined that these events were extragalactic in origin, and that they could be characterized by brief (10$^{-2}$-10$^3$ s) ``prompt" emission in the gamma-ray regime followed by fading ``afterglow" emission in the X-ray, optical, and radio regimes. Examination of the BATSE catalog also led to our current duration-based classification system for GRBs: short GRBs (SGRBs) with burst durations of $<$2 s, and long GRBs (LGRBs) with burst durations of $>$2 s (Kouveliotou et al.\ 1993). In recent years, several observatories dedicated to the study of GRBs, such as the High Energy Transient Explorer 2 (HETE-2; Ricker 1997), the {\it Swift} Gamma-Ray Burst Mission (Gehrels et al.\ 2004), and the {\it Fermi} Gamma-ray Space Telescope (Atwood et al.\ 2009), have been used to study the high-energy properties and afterglow emission of these fleeting events.

The current predominant belief is that two distinct progenitor scenarios can be associated with most GRBs, split according to their duration classifications. The progenitors of SGRBs are still a matter of debate, and have not been observationally identified. The most commonly-cited SGRB progenitor scenario is that of a coalescing compact object binary (consisting of two neutron stars, or a neutron star and a black hole; e.g. Eichler et al.\ 1989, Paczynski 1991, Narayan et al.\ 1992, Gehrels et al.\ 2005), although additional progenitor scenarios such as magnetar formation (Levan et al.\ 2006; Metzger et al.\ 2008, 2011) and accretion-induced neutron star collapse (Qin et al.\ 1998) could also explain some fraction of the SGRB population. LGRBs, however, have been robustly identified as the core-collapse deaths of young and extreme massive stars through observational identifications of supernovae associated with the most nearby events (e.g. Galama et al.\ 1998, Bloom et al.\ 2002a, Hjorth et al.\ 2003, Zeh et al.\ 2004, Woosley \& Bloom 2006, Cobb et al.\ 2010, Berger et al.\ 2011). The most frequently-cited central-engine model for these stellar deaths is referred to as the {\it collapsar} model, where the rapidly-rotating core of a massive star collapses to form a black hole and the progenitor remnants then spiral in and ignite extremely energetic relativistic jets perpendicular to the rotational plane of the black hole that are the source of LGRBs (see Woosley et al.\ 1993, Woosley \& Bloom 2006). However, it is important to note that alternative central-engine scenarios such as magnetar formation are also well-supported by models and evolution of LGRB X-ray light curves (see Troja et al.\ 2007, Lyons et al.\ 2010, Metzger et al.\ 2011).

The spectroscopically-identified supernova counterparts of LGRBs have all been identified as broad-lined Type Ic supernovae SNe. These supernovae are thought to be caused by the collapse of massive stars that have shed their outer hydrogen and helium envelopes (see Filippenko 1997), and the unusually broad absorption features present in the SNe spectra indicate the presence of large ejecta velocities ($\sim$30,000 km s$^{-1}$) due to the effects of velocity broadening (e.g. Galama et al.\ 1998, Patat et al.\ 2001, Pian et al.\ 2006, Modjaz et al.\ 2008).  Unfortunately, more detailed progenitor studies of LGRBs (as well as SGRBs) are extremely challenging. While work on other transient events such as core-collapse supernovae can sometimes utilize pre-explosion imaging to directly investigate their progenitor objects, the rarity and distance of GRBs makes such techniques impossible. Instead, astronomers must rely on a relatively new field: observing and modeling of GRB host galaxies. By focusing on the birthplaces of these events, we can construct a detailed profile of their natal environments and parent stellar populations, placing important constraints on the ages and pre-GRB evolution of progenitor models.

In addition to studying GRB progenitors, dedicated studies of GRB hosts can reveal the population of galaxies that these events are sampling throughout the universe. This is particularly valuable in the case of LGRBs. The short lifetimes of their massive progenitors ($\le$10 Myr; Woosley et al.\ 2002) and their detections out to extremely high redshifts (as high as z$\sim$9.4; Cucchiara et al.\ 2011) have led to LGRBs being cited as potentially powerful tracers of star formation (e.g. Bloom et al.\ 2002b, Chary et al.\ 2007, Savaglio et al.\ 2009). Observations of LGRB emission across the electromagnetic spectrum could be used to trace the metallicity and star formation history of the universe as well as the physics and chemistry of the first stars. However, in recent years several studies suggested a possible connection between LGRBs and low-metallicity galaxies that could threaten their utility as unbiased tracers of star formation (e.g. Stanek et al.\ 2006, Kewley et al.\ 2007, Modjaz et al.\ 2008). Surveys of LGRB host galaxies will allow us to better understand these events' potential environmental biases. With this information we can then effectively employ LGRBs as high-redshift probes to measure, for example, the faint end of the high-redshift galaxy luminosity function (e.g. Robertson \& Ellis 2012, Trenti et al.\ 2012).

In this review I present recent research on the host galaxies of LGRBs. Early studies of these events focused on probing the star-forming nature and metallicity of LGRB host environments, investigating the implications of stellar evolutionary theory that suggested a low-metallicity restriction on LGRB production (Section 2). However, in the past few years several dedicated LGRB host surveys have challenged this picture of progenitor evolution, revealing that LGRBs do indeed sample high-metallicity host environments and emphasizing the potential bias of optically-selected samples that do not properly accommodate ``dark" GRBs (Section 3). Most of these surveys are limited to global studies of host properties due to these galaxies faint and distant nature, but several observations have specifically focused on the most nearby spatially-resolved hosts and what these galaxies can tell us about the immediate local explosion environment of LGRBs (Section 4). At the same time studies of LGRB hosts at higher redshifts have highlighted the great importance of LGRBs as cosmological tools, probing intervening absorbers, the evolution of the mass-metallicity relation, and the earliest star-forming galaxies at $z \sim$ 6-8 (Section 5). Combined, this current  work on LGRB host galaxies is laying important groundwork for future pursuits, which will be able to fully utilize the capabilities of the next generation of telescopes and employ LGRBs as valuable probes of stellar evolution, high-energy phenomena, and the nature of our universe (Section 6).

\section{Early LGRB Host Studies}
Early work on the host galaxies of LGRBs was driven by efforts to explain the origin of these events. Once their association with massive stars was presented, there was immediate interest in the potential utility of these events as tracers of star formation at high redshifts. In addition to their confirmed association with core-collapse supernovae, their connection with star formation was further supported by Bloom et al.\ (2002b), who examined the offsets of LGRBs from their host nuclei and concluded that this distribution was consistent with a UV-tracing (and therefore young massive star tracing) population. At higher redshifts ($z > 4$) Chary et al.\ (2007) found that the cosmic star formation rate density inferred by three LGRBs showed good agreement with the extinction corrected density estimated from Lyman break galaxies. This use of LGRBs as valuable tracers of early universe star formation continues today; for more discussion see Section 5.2.

Fruchter et al.\ (2006) compared the morphologies of LGRB and core-collapse supernova host galaxies, finding that the LGRB environments were significantly different than those of core-collapse supernovae out to $z \sim 1$. LGRB host galaxies as a population were fainter and more irregular than core-collapse hosts, and within their hosts LGRBs were more likely to be localized in the brightest UV regions of the galaxy, which are associated with concentrated populations of young massive stars. Similarly, Svensson et al.\ (2010) found that LGRBs tend to occur in smaller lower-mass galaxies than core-collapse supernovae and are highly concentrated in regions of higher absolute surface luminosity. Leloudas et al.\ (2010) applied a similar analysis to the locations of Wolf-Rayet stars in M83 and NGC 1313 and similarly conclude that these presumed LGRB progenitors, particularly the carbon-rich subtype, trace host light in a similar manner to the Type Ic SNe associated with LGRBs (although it is important to note that Wolf-Rayet stars have not been observationally associated with either Type Ic SNe or LGRBs; see below).

Stanek et al.\ (2006) found that the metallicities of five nearby ($z < 0.3$) LGRB hosts were lower than their equally-luminous counterparts in the local star-forming galaxy population, placing them below the standard L-Z relation for star-forming galaxies (where galaxies with higher masses, and therefore luminosities, are generally found to have higher metallicities, e.g. Lequeux et al.\ 1979, Skillman et al.\ 1989, Zaritsky et al.\ 1994, Tremonti et al.\ 2004). Based on these results, Stanek et al.\ (2006) proposed that LGRB formation was limited by a strong metallicity threshold. Wolf \& Podsiadlowski (2007) also suggested a metallicity-dependent efficiency decrease or cut-off for LGRB formation of log(O/H) + 12 $\sim$ 8.7, finding that LGRB hosts are under-luminous (and presumably also low-metallicity) relative to their star formation rates.

Following these results, Modjaz et al.\ (2008) demonstrated that nearby LGRB host galaxies had systematically lower metallicities than the host galaxies of nearby ($z < 0.14$) broad-lined Type Ic SNe (the same supernova subtype sometimes observed in conjunction with LGRBs), falling on opposite sides of a narrow metallicity threshold. From these results Kocevski et al.\ (2009) again proposed a metallicity-dependent cut-off or decrease in LGRB formation efficiency of log(O/H) + 12 $\sim$ 8.7. Combined, this work on the early sample of LGRB hosts all suggested that these events may be restricted to galaxies with lower metallicities, introducing a potential bias that could threaten their utility as effective star formation tracers. However, the nature of this apparent bias and its potential connection to external factors (such as the dominance of lower-mass galaxies in the local universe or the sample effects of undetected dust-obscured LGRBs) was a matter of much debate and is discussed in detail later in this review.

A metallicity bias for LGRB progenitors was supported by stellar evolutionary theory. Due to the association of LGRBs with stripped-envelope supernovae, carbon- and oxygen-rich Wolf-Rayet stars are frequently proposed as the most likely progenitor candidates (e.g., Hirschi et al.\ 2005, Yoon et al.\ 2006, Langer \& Norman 2006, Woosley \& Heger 2006). Wolf-Rayet star winds are driven by radiation pressure on spectral lines, leading to a dependence of the wind-driven mass loss rate on surface metallicity (e.g. Vink et al.\ 2001, Kudritzki 2002, Vink \& de Koter 2005).  As a result, surface rotation velocities for Wolf-Rayet stars decrease at higher stellar metallicites, a consequence of the higher mass loss rate (Kudritzki \& Puls 2000, Meynet \& Maeder 2005). For young massive stars like these, the metallicities of their host environments can be adopted as the natal metallicities of the stars themselves. This therefore implies that the higher wind-driven mass loss rates in metal-rich environments would rob these massive stars of too much angular momentum, preventing them from rotating rapidly enough to produce a LGRB and suggesting that LGRBs should either be restricted to low-metallicity environments (Yoon et al.\ 2006, Woosley \& Heger 2006) or produce weaker explosion energies at higher metallicities (MacFadyen \& Woosley 1999).

However, it is important to note that no stripped-envelope supernova progenitors have ever been confidently identified and classified (e.g. Gal-Yam et al.\ 2005, Smartt 2009,Van Dyk 2011, Corsi et al.\ 2012), and Wolf-Rayet stars have never been observationally confirmed as supernova progenitors of any sort (however, see Cao et al.\ 2013 and Groh et al.\ 2013 for recent discussion of a candidate Wolf-Rayet progenitor for the Type Ib supernova iPT13bvn). Furthermore, observations of Local Group massive star populations have revealed that the Wolf-Rayet population actually {\it decreases} strongly at lower metallicities, particularly the carbon- and oxygen-rich subtypes (Massey 2003), suggesting that these proposed progenitors may be extremely rare in LGRB host environments. Our assumptions of metallicity effects on progenitor evolution are also dependent upon a single-star core-collapse evolutionary model for LGRB progenitors. Evolution in an interacting binary, for example, could allow a metal-rich LGRB progenitor to maintain a sufficiently high rotation rate despite stronger line-driven winds (e.g. Podsiadlowski et al.\ 2004, 2010; Fryer \& Heger 2005; van den Heuvel \& Yoon 2007; de Mink et al.\ 2013).

It has also been suggested that the proposed connection between LGRBs and metal-poor environments may merely be an artifact of other key environmental properties favored by LGRBs. LGRBs are expected to occur in galaxies with younger stellar populations, which typically have lower metallicites as well (Bloom et al.\ 2002b, Berger et al.\ 2007). LGRBs also occur in galaxies with high specific star formation rates (Stanek et al.\ 2006; Castro Ceron et al.\ 2006, 2010; Savaglio et al.\ 2009). Recent results suggesting a relation between star formation and metallicity have proposed that the apparent metallicity trend in LGRBs is merely an artifact of hosts with high specific star formation rate (Mannucci et al.\ 2011), although this proposal is at odds with a number of more recent surveys (see further discussion in Section 3.3). It is also possible that LGRBs are occurring in low-metallicity regions within galaxies that have apparent high global metallicities, a consequence of metallicity inhomogeities (Niino 2011). Finally, even if LGRBs do show a bias toward low-metallicity environments, the metallicity evolution of the universe yields lower mean metallicites at $z > 1$, which could well make LGRBs excellent tracers of star formation at higher redshifts (Fynbo et al.\ 2006, Berger et al.\ 2007, Kocevski et al.\ 2009).

Much of this early host metallicity work was based upon a small sample of extremely nearby ($z \lesssim 0.3$) and atypical LGRBs: GRBs 980425, 020903, 030329, 031203, and 060218. With the exception of GRB 030329, these bursts have considerably lower energies and luminosities than the general LGRB population, marking them as ``subluminous" bursts. They also show a very low degree of collimation, with quasi-spherical geometries rather than the narrow opening angles inferred for most LGRBs (e.g. Sari et al.\ 1999, Frail et al.\ 2001; see also the sample of Levesque et al.\ 2010c). Finally, the relative rate of these sub-luminous bursts is much higher than that of the LGRB population as a whole. Combined, these properties suggests that such bursts could potentially represent a phenomenologically-distinct subclass (e.g. Kulkarni et al.\ 1998, Soderberg et al.\ 2006, Cobb et al.\ 2006, Zhang et al.\ 2012b; see also Section 4). To better understand whether LGRBs can effectively sample the general star-forming galaxy population, we require larger samples of galaxies that have hosted ``typical" (referred to by Stanek et al.\ 2006 as ``cosmological") LGRBs; that is, LGRBs with luminosities that are detectable in the more distant universe where these events are proposed as useful probes. As a result, more recent work has focused on dedicated observational surveys of LGRB host galaxies.

\section{LGRB Host Surveys in the {\it Swift} Era}
The launch of the {\it Swift} Gamma-Ray Burst Mission (Gehrels et al.\ 2004) has had a tremendous impact on the study of GRBs. The sensitivity of the Burst Alert Telescope (BAT) to high-redshift events has allowed for a tremendous increase in the total sample of GRB detections, including the record-setting GRBs 090423 and 090429B at $z\sim$8-9 (e.g. Salvaterra et al.\ 2009, Tanvir et al.\ 2009, Cucchiara et al.\ 2001). Furthermore, the follow-up capabilities with the X-Ray Telescope (XRT) and Ultraviolet/Optical Telescope (UVOT) have made it possible, through afterglow detections, to obtain precise positions of (almost) all GRBs detected by {\it Swift}. Such detections and positions are crucial for the identification of LGRB host candidates and the assembly of large host galaxy samples. Results from {\it Swift} have made it possible to expand LGRB host surveys from the small handful of well-studied hosts described above to samples that number in the dozens and employ a variety of different selection criteria, allowing for statistically robust analyses and a diverse sample that can more effectively represent the full range of progenitor environments. Below we discuss the largest and most recent surveys made possible by the results of {\it Swift}.

\subsection{Metallicity and LGRB Production}
Levesque et al.\ (2010a,b) conducted a large-scale spectroscopic survey of $z < 1$ LGRB host galaxies, using the Keck telescopes at Mauna Kea Observatory and the Magellan telescopes at Las Campanas Observatory. Restricted to confirmed LGRBs with confidently-associated and observable host galaxies, the survey obtained rest-frame optical spectrophotometry of 13 LGRB host galaxies and supplemented these results with four additional hosts that had high-quality spectroscopic data available in the literature: GRB 980425 (Christensen et al.\ 2008), GRB 990712 (K\"{u}pc\"{u} Yoldas et al.\ 2006), GRB 030528 (Rau et al.\ 2005), and GRB 050824 (Sollerman et al.\ 2007). Combined with multi-band host photometry from Savaglio et al.\ (2009), these spectra were used to determine a number of key parameters for the LGRB hosts, including metallicity, ionization parameter, young stellar population age, SFR, and stellar mass. Levesque et al.\ (2010a) revealed that most of these LGRB host galaxies fell below the general L-Z relation for star-forming galaxies and were statistically distinct from the host galaxies of stripped-envelope core-collapse supernovae as well as the larger star-forming galaxy population.

Following this, Levesque et al.\ (2010b) determined stellar masses for these LGRBs hosts and compared them to star-forming galaxies on a mass-metallicity diagram (see Figure 1). These results revealed that LGRB hosts followed their own statistically robust mass-metallicity relation out to $z \sim 1$ that is offset from the general mass-metallicity relation for star-forming galaxies by an average of $-0.42\pm 0.18$ dex in metallicity. This marks LGRB hosts as distinct from the host galaxies of Type Ibc supernovae; these events do appear to more effectively trace the star-forming galaxy population, although the nature of a potential low-metallicity bias among stripped-envelope supernova subtypes is still a matter of debate. (e.g. Modjaz et al.\ 2008, 2011; Kelly \& Kirshner 2012; Sanders et al.\ 2012).

Most surprisingly, the sample included several LGRB hosts that fell above the predicted cut-off metallicities for LGRB formation (see also Elliott et al.\ 2013). This proved that, contrary to previous predictions, LGRB progenitors can indeed be produced in metal-rich environments (for more discussion see Section 4). In the absence of a production cut-off, single-star evolutionary theory predicts that LGRBs at high metallicity can indeed exist, but are expected to produce less energetic explosions (e.g. MacFadyen \& Woosley 1999). Stanek et al.\ (2006) found evidence of this correlation in their sample of five nearby LGRB hosts. However, Levesque et al.\ (2010c) extended this to a larger survey sample and found no correlation between metallicity and the isotropic or beaming-corrected energy release of GRBs in the gamma-ray regime ($E_{\gamma,iso}$ and E$_{\gamma}$, respectively; see Figure 2). This lack of a correlation appears to demonstrate that, while there is at least some apparent connection between host metallicity and the rate of successful LGRBs, it has no clear impact on their detailed explosive properties (although it is worth noting that the full parameter space of GRB prompt emission properties has not yet been examined). This lack of a strong connection between LGRB properties and metallicity is surprising considering the critical role that metallicity plays in massive star evolution, and could suggest that alternate progenitor scenarios beyond the single-star collapsar model, such as binary evolution channels or magnetar formation, play an important role in LGRB production. Alternately, it illuminates flaws with the previous general of single-star stellar evolutionary models, particularly at low metallicity. The most recent generations of these models include detailed treatments of stellar rotation that can decouple surface mass loss effects from the evolution of the stellar interior, leading to a disconnect between natal metallicity and final explosive properties for core-collapse events (see, for example, Ekstr\"{o}m et al.\ 2011; Georgy et al.\ 2012, 2013; Groh et al.\ 2013).

\subsection{The Hosts of ``Dark" GRBs}
One shortcoming of many LGRB host studies is their reliance of the optical afterglows of LGRBs, which are typically used for precise LGRB astrometry and the subsequent association of these events with their hosts. However, only about half of the GRBs detected by {\it Swift} have an associated optical detection (Perley et al.\ 2009), compared with a nearly 100\% detection rate for X-ray afterglows (Gehrels 2008). While a number of these non-detections can be attributed to observational difficulties, in some cases the optical afterglow of an LGRB is found to be absent to below very stringent limits. As a result, objects such as these, commonly referred to as ``dark" GRBs, are typically excluded from host studies.

Perley et al.\ (2009) used the imaging capabilities of Keck to specifically detect and study the host galaxies of 14 ``dark" {\it Swift} LGRBs. From this work they found evidence for large amounts of line-of-sight dust (median of $A_V \sim$ 0.5 mag) toward these dark bursts (with several as high as $A_V \gtrsim 2.5$ mag) and concluded that the phenomenon of dark LGRBs could likely be attributed to extinction effects. However, while the line-of-sight extinction of the dark LGRBs is generally high, they note that the nature of this extinguishing dust is still unclear. Given the blue color of some of the Perley et al.\ (2009) hosts and assuming the dust is widespread, the whole galaxy should show a gray extinction curve (see also Perley et al.\ 2008).  However, other afterglows that are heavily dust-extinguished do show substantial non-``gray" reddening (e.g. Greiner et al.\ 2011, Kr\"{u}hler et al.\ 2012, Perley et al.\ 2013), suggesting that this trait is at least not universal for LGRBs. An alternate explanation is that the dust obscuring the optical afterglows in Perley et al.\ (2009) could be patchy, localized around the progenitor while still distant enough to avoid being destroyed by the burst itself (Waxman \& Draine 2000).

Kr\"{u}hler et al.\ (2011) observed a sample of 8 dark GRB host galaxies, selecting GRB afterglows with $A_V > 1$ and performing dedicated follow-up searches for their hosts. While this small dark GRB host sample was found to overlap with the population of optically-bright LGRB hosts, in general dark burst hosts have higher luminosities and metallicities, suggesting that the large line-of-sight $A_V$s preferentially select more massive, and hence higher-metallicity, host galaxies than their optically-bright counterparts. Similarly, Perley et al.\ (2013) targeted the hosts of 23 dark GRBs (events with $A_V > 1$ afterglows) and find substantial differences between the hosts of dust-obscured and optically-bright GRBs. Dark GRB hosts are an order of magnitude more massive that optical-bright GRB hosts within the same redshift range, and also show evidence of higher luminosities, redder colors, higher global dust content, and older stellar populations (although interestingly, their star formation rates do not appear to be atypically extreme; see Perley \& Perley 2013). A number of individual dark GRBs have also been observed in high-metallicity environments. These include GRBs 020819 and 051022, two of the highest-metallicity LGRB hosts in Levesque et al.\ (2010b; see also Graham et al.\ 2009, Levesque et al.\ 2010d), and GRB 080607, a $z = 3.036$ dark GRB with a solar metallicity inferred from afterglow absorption features (Prochaska et al.\ 2009). Combined, these results have important implications for the potential selection effects present in LGRB host surveys based only on optical afterglows, chief among them a possible bias towards lower-metallicity LGRB hosts that may not be representative of the general population and a less massive population that does not effectively sample the high-mass end of the mass-metallicity relation.

Infrared and radio detections of LGRB afterglows offer a powerful alternative means of selecting events for host galaxy follow-up that does not select against dark GRBs. Zauderer et al.\ (2013) examined the environments of two highly-extincted LGRB host galaxies, using radio observations to localize the LGRBs to their hosts. By probing the circumburst environments of these events, they infer that the inferred mass loss rates are larger by about an order of magnitude compared to optically-bright bursts. This suggests that dark GRBs, which as previously noted occur in higher-metallicity environments on average, may have stronger line-driven winds, a conclusion in good agreement with the predictions of single-star evolutionary theory where massive stars are expected to have a mass loss-metallicity relation of $\dot M_w \propto Z^{0.7}$ (Vink et al.\ 2011; see also Section 1), and with the prediction of Levesque et al.\ (2010d) who proposed a potential correlation between dark bursts, high-metallicity environments, and large amounts of circumburst extinction produce by progenitor mass loss. 

\subsection{LGRB Host Studies Beyond the Optical}
Studies of LGRB host properties have also extended to longer wavelengths, with several sub-mm and radio searches for continuum emission from host galaxies. Interpretations of the detections, upper limits, and SEDs from these surveys have varied, but do support the conclusion that LGRB host galaxies are not representative of the general galaxy population (e.g. Berger et al.\ 2003, Tanvir et al.\ 2005, Le Floc'h et al.\ 2006, Michalowski et al.\ 2008). Several of these studies have specifically investigated star-formation behavior in these hosts; Berger et al.\ (2003) reported immense star formation rates in excess of several hundred M$_{\odot}$/yr for some of their LGRB hosts, while radio-derived SFRs from Stanway et al.\ (2010 and Hatsukade et al.\ (2012) imply that the galaxies either have very little dust-obscured star formation or starbursts that are too young to establish radio continuum emission. While sub-mm and radio studies of LGRB hosts remain challenging, Wang et al.\ (2012) note that longer-wavelength studies could potentially be extremely powerful at very high redshifts ($z\gtrsim20$), offering the possibility of metal line detections that sample the heavy elements produced by the first supernovae. However, despite being based on sub-mm and radio observations, several of these studies also specifically address the challenges of optically-biased sample selection, particularly in the pre-{\it Swift} sample; Le Floc'h et al.\ (2006) note that early identification techniques could have biased their sample against dusty galaxies. 

The Optically Unbiased GRB Host (TOUGH) survey has searched for the hosts of 69 LGRBs detected by {\it Swift}, selected based on the presence of X-ray afterglows rather than optical afterglows or previously-detected hosts and thus eliminating any substantial optical bias. Observations of these LGRB hosts include optical and infrared photometry from the Very Large Telescope (VLT) as well as radio data (Hjorth et al.\ 2012, Michalowski et al.\ 2012). So far, 53 of these hosts have known redshifts, with an average redshift of $z = 2.23$ that is higher than previous host surveys, particularly those based on primarily on pre-Swift data (Jakobsson et al.\ 2012). These include ground-based emission line redshifts of galaxies at $z \sim 2$ using the X-shooter spectrograph at the VLT (Kr\"{u}hler et al.\ 2012); these same spectra the possibility of future determinations of galaxy properties such as metallicity in this larger and more distant sample that can be directly compared with the Levesque et al.\ (2010a,b) results. Similar to Perley et al.\ (2013), the results of Hjorth et al.\ (2012) are consistent with the idea that dark LGRBs have more massive hosts than LGRBs with optical afterglows. Michalowski et al.\ (2012) state that the star formation rates and dust attenuation of the $z < 1$ TOUGH population as a whole are found to be consistent with the general star-forming galaxy population; however, it is worth noting that their analyses of these parameters typically result in broad ranges or limits that are not strongly constraining and do not rule out a scenario in which LGRBs prefer sub-luminous galaxies. Finally, Milvang-Jensen et al.\ (2012) find that, contrary to previous predictions, GRB host galaxies are not, uniformly, Lyman $\alpha$ emitters (as would be expected given their anticipated star formation rates), and suggest that the absence of detected Lyman $\alpha$ emission for many of these hosts can be at least in part attributed to a high dust content, similar to the proposed explanation for dark LGRBs. 

\subsection{Implications of LGRB Host Surveys} 
\subsubsection{LGRBs and Star Formation}
While the sampling and observational techniques of these recent surveys differ, all share the common conclusion that LGRBs are not exclusively formed in low-metallicity environments. These results have powerful implications for future progenitor models and the role that metallicity plays in LGRB production and progenitor evolution. The optically-selected LGRB host galaxies from Levesque et al.\ (2010a,b) appear to occur preferentially in low-metallicity environments, but there is no evidence of a cut-off metallicity above which LGRB production is suppressed and no correlation between the metallicity of the progenitor environments and the gamma-ray energy release produced by the accompanying LGRB. While the inclusion of dark GRB hosts appears to increase the average stellar mass and metallicity of the overall LGRB host sample, Perley et al.\ (2013) note that this still does not result in a sample that represents the general star-forming galaxy population (although this agreement with the general population does improve at $z > 1.5$, a possible effect of redshift-dependent metallicity evolution; see also Section 5.1). They conclude that the LGRB rate is indeed dependent on host metallicity, but that this dependence is best represented by a gradual trend rather than a strict metallicity cut-off, in agreement with Levesque et al.\ (2010b). This is also supported by Zauderer et al.\ (2013), with high-mass-loss progenitors producing dark GRBs in presumably-high-metallicity environments. 

There is ongoing debate concerning the connection between LGRBs and star formation, and how any apparent metallicity trends impact the use of these events as effective tracers. Recently, Mannucci et al.\ (2011) proposed that LGRBs purely trace star formation rather than any kind of metallicity preference. They contend that LGRBs only show a metallicity bias as a consequence of the ``fundamental metallicity relation" (FMR) that is proposed to connect metallicity with both stellar mass {\it and} specific star formation rate for star-forming galaxies (Mannucci et al.\ 2010). One consequence of this would be an LGRB host mass-metallicity relation that lies above the Lyman-break galaxy relation; however, the opposite is observed (Laskar et al.\ 2011; for more discussion see Section 5.1). Several other LGRB host studies also disagree with the conclusions of Mannucci et al.\ (2011). Kocevski \& West (2011) find that the correlation between specific star formation rate and metallicity proposed by the FMR is weak and insufficient to fully account for the low-metallicity trend seen in LGRB hosts. Graham \& Fruchter (2013) considered the predicted LGRB rate and distribution that would be produced by a purely SFR-based selection effect, and found clear inconsistencies with observed $z < 1$ LGRB rates that could only be rectified by including a metallicity dependence or other similar effects. Finally, Perley et al.\ (2013) find that, even with the inclusion of the dark GRB host sample, the LGRB rate is a strong function of the host environment properties. Collectively, these results suggest that, while LGRBs are certainly strongly associated with star formation, their rates and host galaxies demand additional complicating factors. Using LGRB hosts as a test sample, these results also call into question the robustness of the proposed FMR.

\subsubsection{Progenitor Models of LGRBs}
The results of Zauderer et al.\ (2013), associating LGRBs with high-mass-loss progenitors, come with the surprising implication that massive stars can lose a great deal of mass without a corresponding loss of angular momentum in the core. Along with observations of LGRBs in high-metallicity explosion environments, such findings are at odds with the standard predictions of stellar evolutionary theory discussion in Section 1. These predict that at high metallicities, high mass loss rates will decrease the surface rotation velocities of massive stars, robbing them of angular momentum and preventing the core from sustaining the rapid rotation rate required for LGRB production (Kudritzki \& Puls 2000, Meynet \& Maeder 2005, Yoon et al.\ 2006, Woosley \& Heger 2006). However, alternative evolutionary scenarios could make it possible for LGRB progenitors to lose a great deal of mass without sacrificing angular momentum. Recent models of stellar evolution with rotation consider the complex connection between {\it surface} angular momentum loss and {\it core} angular momentum loss, finding that single-star progenitors are still capable of producing LGRBs across a broad range of metallicities (see Ekstr\"{o}m et al.\ 2012, Georgy et al.\ 2012, Groh et al.\ 2013). Single-star LGRB progenitor models at high metallicities should also consider the effects of anisotropies in stellar winds, a potential explanation for dark GRBs proposed in Levesque et al.\ (2010d). Polar mass ejections remove much less angular momentum than equatorial ejections (Maeder 2002), which would simultaneously increase circumburst extinction along the afterglow line-of-sight {\it and} allow the progenitor to sustain a high rotation rate. Alternately, episodic explosive or eruptive mass loss behavior (such as that observed in massive stars such as luminous blue variables and red supergiants) offers another scenario that allows stars to shed their outer envelopes while retaining their angular momentum.

Moving beyond single-star models, binary progenitor progenitor scenarios are extremely effective at explaining these results. One common binary progenitor model considers the coalescence of two stellar cores - either two massive stars or a massive star and low-mass companion during a  terminal common-envelope phase that ultimately ejects the outer envelope (Fryer \& Heger 2005, Podsiadlowski et al.\ 2010). This type of common-envelope evolution for binaries is predicted to be more common in, but not exclusive to, metal-poor environments, with weaker stellar winds allowing binaries to evolve at closer proximities (Podsiadlowski et al.\ 2004). Another binary model predicts an interim common envelope phase where the outer envelope of the progenitor is ejected or stripped, followed by a contact phase that permits the progenitor to retain angular momentum. This scenario should also occur at a higher rate in metal-poor environments due to a wider range of Roche lobe radii that permit binaries to enter and survive this temporary common envelope phase (Linden et al.\ 2010). Both binary scenarios can also produce high mass loss rates without a significant loss of angular momentum in the progenitor core. Additional production channels such as magnetar formation, a broader range of initial conditions for the traditional collapsar model, or a combination of multiple progenitor channels have all been presented as ways of accommodating these new and evolving properties of LGRB host environments (Metzger et al.\ 2011, Dessart et al.\ 2012).

\section{Local Environments of LGRBs}
Studies of GRB host galaxies are generally restricted to observations of their global morphologies and interstellar medium (ISM) properties due to their prohibitive distances and small angular sizes. Most LGRB hosts are far too distant to discern spatially-resolved host properties such as metallicity and star formation rate gradients. These limitations are important to consider when extrapolating progenitor properties from LGRB host surveys, since it is possible that the global galaxy properties may not be an accurate proxy for the properties of the local LGRB explosion environment. However, there are a handful of notable exceptions among the most nearby and most massive LGRB hosts, all of which have angular sizes large enough to permit spatially-resolved observations. The detailed study of these galaxies offers a unique opportunity to probe the precise natal environments of LGRB progenitors and consider whether global ISM properties are in fact a useful proxy for local conditions. 

At $z = 0.008$, GRB 980425 is the closest LGRB ever observed. Christensen et al.\ (2008) obtained detailed integral field unit spectroscopy of the dwarf SBc spiral host galaxy, finding that the metallicity and star formation rate of the GRB explosion site is comparable to other HII regions of the host, with all metallicities in the host agreeing to within 3$\sigma$ and corresponding to sub-solar metallicities.  Th\"{o}ne et al.\ (2008) used long-slit spectroscopy to probe the $z = 0.089$ Sbc spiral host of GRB 060505, also finding a minimal metallicity gradient within the host galaxy and agreement between the GRB site and other host regions that was within the errors of the metallicity diagnostics (see also Kewley \& Ellison 2008). Le Floc'h et al.\ (2012) used infrared spetro-imaging to examine the close environment of the GRB 980425 Wolf-Rayet region, which is offset from the GRB site but shows evidence of strong and recent ($<$5 Myr) star formation.

More recently, Levesque et al.\ (2010d) measured super-solar metallicities at both the nucleus and explosion site HII region of the $z = 0.41$ GRB 020819 host galaxy, marking the first unambiguous association of an LGRB with a high-metallicity explosion site. Spatially-resolved spectroscopy of the GRB 100316D host complex revealed that the LGRB was concentrated near the lowest-metallicity and most strongly star-forming region of the host (Levesque et al.\ 2011; see Figure 3), although the metallicity gradient of the host was found to be extremely small. Spectra of several sites within the GRB 120422A led to similar results, with a weak metallicity gradient and a nominally lower abundance at the explosion site (Levesque et al.\ 2012a).

These same studies compared the global and local metallicites measured for the small sample of five LGRB hosts where such precision is possible, and concluded that global host properties (such as those determined in smaller or more distant galaxies) can indeed be used as effective proxies for LGRB environments, with global and local metallicities agreeing to within 0.1 dex (although the local metallicities do uniformly fall $\sim$0.1 dex below the global metallicities in each case; see Levesque et al.\ 2011 for further discussion). These results are at odds with the results for stripped-envelope CCSNe from Modjaz et al.\ (2011), who conclude that global oxygen abundances are a very poor proxy for the local host environments, and suggests further evidence of the clear difference between the two events and their progenitor classes. However, it is important to note that the GRBs in these samples are {\it not} necessarily representative of the larger LGRB population. The current picture of LGRBs includes evidence for multiple potential subclasses. GRB 020819 is a member of the ``dark" GRB class described in Section 3.2, lacking an optical afterglow but well-localized within its host due to a well-detected radio afterglow (Jakobsson et al.\ 2005). GRBs 980425, 060505, 100316D, and 120422A are all members of the ``subluminous" class described in Section 2. Three of these GRBs (980425, 100316D, and 120422A) are associated with spectroscopically-confirmed broad-lined Type Ic supernovae; however, GRB 060505 represents a rare and strange case. While a burst duration of 4 s classifies this event as an LGRB, detailed observations found no evidence of any associated supernova despite its proximity. A similar result was found for the 102 s GRB 060614; combined, these two events represent an unusual class of LGRBs that clearly lack the accompanying supernova signature expected for a core-collapse event (Fynbo et al.\ 2006, Della Valle et al.\ 2006, Gal-Yam et al.\ 2006). The classification of these two unusual GRBs and their connection to the broader LGRB (or SGRB) sample is still unresolved (e.g. Ofek et al.\ 2007, Levesque \& Kewley 2007, Zhang et al.\ 2007), and further highlights the possibility of multiple phenomenological subclasses among LGRBs. However, Levesque et al.\ (2012a) did confirm that there does not appear to be any statistically significant difference in the host galaxy metallicities sampled by these different subclasses at $z < 1$.

Collectively, these spatially-resolved studies of nearby LGRB hosts are able to probe the specific natal environments and explosion sites of LGRB progenitors. However, a larger sample of similar observations is necessary before any definitive conclusions can be made or before these results can effectively be generalized to the larger and higher-redshift LGRB host population.

\section{High-Redshift LGRBs} 
\subsection{LGRB Environments from $z > 1$ Afterglow Spectroscopy}
At $z \gtrsim 1.5$, detailed studies of LGRB host environments have thusfar depended on the events' afterglow spectra and observations of rest-frame ultraviolet absorption features. Rapid spectroscopic follow-up of these events make it possible to probe local LGRB host environments ranging from $1.5 \lesssim z \lesssim 4$. From X-ray and optical afterglow spectra we can measure dust and metal column densities in LGRB hosts, as in the recent survey by Schady et al.\ (2010). Afterglow spectra can also be used to determine metallicities, column densities, dust contents, and detailed information about the local ISM and circumburst environments of the events' progenitors, and shed light on intervening systems along the line of sight (e.g. Watson et al.\ 2006, Prochaska et al.\ 2006, El\'{i}asd\'{o}ttir et al.\ 2009, Prochaska et al.\ 2009, Savaglio et al.\ 2012).

Prochaska et al.\ (2007) used afterglow spectra of LGRBs spanning a broad range of redshifts ($1.2 < z < 6.3$) to probe the chemical abundances of the progenitors' nearby ISM environments. From damped Lyman-$\alpha$ (DLA) absorption systems at the GRB redshifts, they determined metallicities spanning from $0.01Z_{\odot} < Z < Z_{\odot}$, ultimately finding that LGRB DLA systems yielded higher metallicity measurements on average that DLAs from QSO spectra. Several other studies have found extremely high metallicities in GRB host environments, relative to both $z < 1$ LGRB samples and other star-forming galaxies at comparable redshifts; Savaglio et al.\ (2012), for example, measure high metallicities for two galaxies at $z \sim 3.57$ in the afterglow spectrum of GRB 090323, and postulate that these galaxies include the LGRB progenitor environment and are in the process of merging. However, Prochaska et al.\ (2007) do note that DLA in LGRB afterglows are expected to probe the denser high-metallicity gas in the inner regions of their host galaxies, while DLAs from QSOs sample the less dense (and less enriched) outer regions of star-forming galaxies. While most LGRB afterglow spectra show evidence of high neutral hydrogen column densities in agreement with this explanation, several exceptions have been noted, including GRB 021004 (Fynbo et al.\ 2005), GRB 060607 (Prochaska et al.\ 2008), and the ambiguously-classified GRB 090426 (Levesque et al.\ 2010e). One proposed explanation for these low neutral hydrogen density measurements in LGRB afterglows is that the hot progenitor's radiation field ionizes most of the nearby neutral hydrogen (Fynbo et al.\ 2005, Watson et al.\ 2013).

More recently, Laskar et al.\ (2011) used afterglow absorption spectroscopy and deep Spitzer imaging of 20 LGRB hosts to study the mass-metallicity relation at $z \sim 3-5$ (see Figure 4). From their observations they find a wide range of metallicities in LGRB host environments. However, the mean metallicity of these environments is lower than measurements at $z < 3$, in agreement with prior evidence of redshift-dependent metallicity evolution (e.g. Shapley et al.\ 2004, Erb et al.\ 2006, Savaglio 2006, Chary et al.\ 2007, Dav\`{e} \& Oppenheimer 2007, Liu et al.\ 2008). Similarly, they find tentative evidence for a mass-metallicity relation in LGRB hosts at $z \sim 3.5$ along with evidence that this relation has evolved to systematically lower metallicites compared to relations at lower redshift ($z \sim 2$). Finally, the mass-metallicity relation that Laskar et al.\ (2011) determine for $z \sim 3-5$ LGRB hosts lies below the relation traced by Lyman-break galaxies.

It is important to note that comparing host environment properties determined from rest-frame optical emission features and rest-frame ultraviolet absorption features is extremely difficult. Even interstellar medium diagnostics within a single wavelength regime require careful calibrations in order to perform effective comparisons (see, for example, Kewley \& Ellison 2008), and no multi-wavelength calibration has yet been established between optical and UV environmental diagnostics. While afterglow absorption metallicities can be effectively considered as a set, comparing these ISM properties to those based on emission line diagnostics for $z < 1$ hosts cannot, at the moment, be done reliably. Such studies are dependent on the development of new stellar population synthesis and photoionization models that can be used to parameterize calibrations of diagnostic calibrations that span multiple wavelength regimes (for more discussion see Levesque et al.\ (2012b).

\subsection{LGRBs in the Early Universe} 
Studies of LGRB hosts have also recently extended out to the extremely high-redshift universe. At $z \sim 8.2$, GRB 090423 has the highest spectroscopically-determined redshift of any LGRB (Salvaterra et al.\ 2009, Tanvir et al.\ 2009), while Cucchiara et al.\ (2011) report a photometric redshift of z $\sim$ 9.4 for GRB 090429B. At these extreme redshifts, LGRB host galaxies can function as valuable cosmological tools for probing early star formation. Tanvir et al.\ (2012) used deep Hubble Space Telescope (HST) imaging to determine limits on the luminosities of six $5 < z < 9.5$ LGRB hosts, finding that all six lie below L*. With the assumption that the LGRB rate is proportional to star formation rate, this suggests a galaxy luminosity function with a steep faint-end slope and decreasing L*. Trenti et al.\ (2012) model a similar galaxy luminosity function and conclude that, if LGRBs extend to faint galaxies in this distribution, detections of LGRB hosts at $z > 6$ should be extremely rare (in agreement with the Tanvir et al.\ 2012 limits) while detections of $z \sim 4-5$ LGRB hosts should help to improve limits on faint galaxy formation. Both Tanvir et al.\ (2012) and Trenti et al.\ (2012) emphasize that future deep imaging studies of high-redshift LGRB hosts will help to further probe the role that the faintest star-forming galaxies play in reionization. 

However, this use of LGRBs as high-redshift tracers of star-formation is also one of the primary motivations driving current work on nearby LGRB host environments and progenitor scenarios. Both Tanvir et al.\ (2012) and Trenti et al.\ (2012) note that these future studies will require a more detailed understanding of LGRB progenitor scenarios and any potential environmental biases that maybe accompany them, although it is true that a low-metallicity preference should become less problematic at higher redshifts due to metallicity evolution effects. Other studies that focus on high-z LGRB rates rather than host properties have also noted the importance of understanding potential LGRB progenitors and environmental biases when considering their utility in the early universe (Kistler et al.\ 2009, Robertson \& Ellis 2012). Interestingly, these same studies have also found an apparent excess of high-redshift LGRBs relative to the rates predicted by current models of star formation in the early universe.

Furthermore, it is possible that the progenitors of high-z LGRBs are distinct from their nearby counterparts. GRB classifications are based on their observer-frame durations, with the phenomenological distinction between burst classes based on their characteristic timescales. It is interesting to note that, with time dilation effects, the {\it rest-frame} duration of, for example, GRB 090423, is only 1.1s (Tanvir et al.\ 2009). However, this is in part a consequence of the {\it Swift} Burst Alert Telescope observing high (MeV) rest-frame photon energies for these events - when considering the the overall lightcurves and high energy releases, these remain consistent with the long-duration class and a core-collapse massive star origin (Zhang et al.\ 2009; see also Amati 2006, Peng et al.\ 2012 for further discussion of the correlation between pulse width and photon energy. The nature of these massive progenitors, however, is expected to evolve as we move from the local sample out to high redshifts. Belczynski et al.\ (2010) found that, independent of metallicity effects, the majority of LGRBs at $6 \lesssim z \lesssim 10$ originate from the Population II stars that dominate this epoch, while LGRBs at $z \lesssim 2$ are expected to originate largely from Population I stars. Population II stars have lower metallicities than their later-universe counterparts, but still sample an era of star formation where the metal abundance is high enough to contribute significantly to cooling effects in collapsing gas clouds, making this distinct from star formation processes in the very early universe. At the same time, approximately $\sim$10\% of the $z \lesssim 6 \lesssim 10$ LGRB population is expected to arise from Population III stars (Campisi et al.\ 2011), the universe's first generation of stars whose unique and extremely low metallicities pose a challenge to evolutionary and core-collapse models (e.g. Hirschi 2007, Ohkubo et al.\ 2009, Yoon et al.\ 2012). Combined, these questions about the nature of early universe LGRB progenitors suggest the need for caution when applying results from nearby LGRB host galaxies to the high-redshift galaxy sample. 

\section{The Future of LGRB Host Studies} 
Recent surveys of LGRB host galaxies have dramatically shaped our understanding of these events' natal environments. LGRBs appear to occur preferentially, but {\it not} exclusively, in low-metallicity environments (Levesque et al.\ 2010a,b; Perley et al.\ 2009, 2013), and show no statistically significant correlation between their gamma-ray energy release and host metallicity (Levesque et al.\ 2010e). While at least some LGRBs have been confidently associated with core-collapse supernovae and massive star progenitors, a number of different pre-explosion evolutionary channels have been proposed, including both binary and single-star models. The nature of LGRB progenitors and the potential connection between their evolution and host environments are both critical to understanding how these events sample the star-forming galaxy population.

Current LGRB host surveys can be extended out to higher redshifts ($z \sim 2$) using a combination of ground-based optical and infrared surveys (such as X-shooter on the VLT; Kruhler et al.\ 2012) and optical and ultraviolet observations with HST. These observations, focused on rest-frame ultraviolet spectroscopy, will also allow for effective direct comparisons with the environmental properties determined from afterglow spectroscopy. Additional spatially-resolved host studies will continue to help us hone in on the precise explosion environments of these events. Finally, dedicated and deep follow-up HST imaging campaigns of high-redshift ($4 \lesssim z \lesssim 10$) LGRBs will increase our number of high-redshift host detections and upper limits, directly testing the effectiveness of LGRBs as high-redshift probes of early star formation. For all of these studies, it will be critical to consider our latest understanding of the differences between dark and non-dark GRB hosts, and select future observational samples accordingly in order to avoid selection biases. \\

The study of gamma-ray burst host galaxies has undergone an extraordinary evolution over the past ten years. In the coming decade, a wealth of new observational facilities will further expand the horizons of this field. New sub-millimeter and radio facilities such as the Cerro Chajnantor Atacama Telescope (CCAT) and the Atacama Large Millimeter Array will expand the studies of these events to new wavelengths. CCAT, for example, will be able to detect star-forming galaxies at extremely high redshifts ($z\gtrsim$6) from the ground due to the shift of these galaxies' spectral energy distribution peaks in the sub-millimeter, making it possible to obtain new host detections for high-redshift LGRBs. Observatories such as the Extremely Large Telescopes (ELTs) will also dramatically extend our reach into both the local and high-redshift universe. The ELTs will allow us to extend extragalactic stellar astronomy - particularly studies of nearby resolved massive star populations  - well beyond the currently-studied galaxies of the Local Group, offering observations of extremely low-metallicity massive stars that can act as valuable local analogs of LGRB progenitors. These facilities will also be capable of much more powerful rapid follow-up of LGRB afterglows, and will be able to acquire host galaxy spectroscopy at greater distances and in unprecedented detail, further expanding studies of LGRB host environments and intervening absorbing systems at $1 \lesssim z \lesssim 6$. Our sample of LGRB hosts will grow dramatically in both size and scope as a result of these telescopes' capabilities, extending out to the redshift regime where metallicity evolution becomes apparent (e.g. Erb et al.\ 2006, Dav\'{e} \& Oppenheimer 2007) and offering an immense new sample that can be used to test the relationship between LGRB progenitors, metallicity, and star formation.

Finally, the James Webb Space Telescope (JWST) will be capable of detecting galaxies out to $z \sim 15$; strongly star-forming galaxies with bright H and He recombination lines and a strong UV continuum are the prime candidates for these detections. The Near Infrared Spectrograph (NIRSpec) and Near Infrared Camera (NIRCam) on JWST should be capable of detecting Ly$\alpha$ and He II 1640\AA\ emission features at $10 \lesssim z \lesssim 15$ through spectroscopy and narrow-band imaging, and bright 1500\AA\ UV continuum emission from these star-forming galaxies should be detectable by NIRCam. In addition, the Mid Infrared Instrument (MIRI) will be capable of detecting strong H$\alpha$ signatures of galaxies at $8 \lesssim z \lesssim 15$. The association of LGRBs with high rates of star formation make the the hosts of these events ideal candidates for JWST's extraordinarily high-redshifts studies of the earliest star-formation in the universe; with JWST it should be possible to obtain confident host detections and associations for LGRBs at $z \gtrsim 9$, giving us a window into the natal environments of Population III stars and their core-collapse deaths.

These new observatories, ranging from the sub-millimeter to the optical, all offer valuable future opportunities for exploring LGRB host galaxies and the capabilities that they hold as potentially powerful probes of our universe. By continuing to make the study of GRBs and their host galaxies a priority, we will ensure that we are equipped with the knowledge that will be required to take full advantage of the frontiers opened up by these facilities and immediately begin making new leaps in our understanding of these enigmatic events.\\

Many thanks to my collaborators in our work on GRB host galaxies, including Megan Bagley, Edo Berger, Joshua Bloom, Ryan Chornock, Andrew Fruchter, John Graham, Lisa Kewley, Tanmoy Laskar, Dan Perley, Alicia Soderberg, and H. Jabran Zahid. Edo Berger provided valuable feedback regarding the content of this manuscript. Finally, I would like to thank the anonymous referee for their useful and insightful suggestions that improved the paper during the review process. Support for this work was provided by NASA through Einstein Postdoctoral Fellowship grant number PF0-110075 awarded by the Chandra X-ray Center, which is operated by the Smithsonian Astrophysical Observatory for NASA under contract NAS8-03060, and through Hubble Fellowship grant  number HST-HF-51324.01-A from the Space Telescope Science Institute, which is operated by the Association of Universities for Research in Astronomy, Incorporated, under NASA contract NAS5-26555.

\clearpage
\begin{deluxetable}{l c c c c c c c c}
\tabletypesize{\scriptsize}
\tablewidth{0pc}
\label{tab:gals} 
\tablecolumns{9}
\tablecaption{ISM Properties of LGRB Host Galaxies}
\tablehead{
\colhead{Host Galaxy}
&\colhead{$z$}
&\multicolumn{3}{c}{log(O/H) + 12\tablenotemark{a}}
&\colhead{E($B-V$)\tablenotemark{b}}
&\colhead{$M_B$ (mag)\tablenotemark{c}}
&\colhead{SFR (M$_{\odot}$/yr)}
&\colhead{log M$*$}\\ \cline{3-5}
\multicolumn{2}{c}{}
&\colhead{$T_e$}
&\colhead{R$_{23}$}
&\colhead{PP04}
&\multicolumn{3}{c}{}
&\colhead{(M$_{\odot}$)}
}
\startdata
GRB 980425 &0.009 &\nodata &$\sim$8.40    &8.28        &0.34  &-17.6 &0.57 &9.22 $\pm$ 0.52 \\ 
GRB 060218 &0.034 &7.62 &8.21   &8.07  &0.01  &-15.9 &0.03 &8.37 $\pm$ 0.14 \\
GRB 100316D\tablenotemark{d} &0.059 &\nodata &\nodata &8.20 &0.12 &-19.6 &1.7 &\nodata \\
GRB 031203\tablenotemark{e} &0.105 &7.96 &8.27          &8.10    &1.17   &-21.0 &4.8 &8.26 $\pm$ 0.45 \\ 
GRB 030329 &0.168 &7.72 &8.13          &8.00      &0.13   &-16.5 &1.2 &7.91$^{+0.12}_{-0.44}$ \\
GRB 020903 &0.251 &\nodata &8.07          &7.98   &0.00      &-18.8  &1.7 &8.79$^{+0.19}_{-0.24}$ \\ 
GRB 120422A\tablenotemark{d} &0.283 &\nodata &8.40 &8.20 &0.31 &-19.4 &2.7 &\nodata \\
GRB 050826 &0.296 &\nodata &8.83 &\nodata &0.48  &-19.7 &4.11 &10.10$^{+0.22}_{-0.26}$ \\ 
GRB 020819\tablenotemark{f} &0.410 & \nodata &9.0 &8.8 &0.71 &\nodata &23.6 &10.65 $\pm$ 0.19 \\ 
GRB 990712 &0.434 &\nodata &$\sim$8.40 &\nodata  &0.57    &-18.6            &10.7\tablenotemark{g} &9.15 $\pm$ 0.04 \\ 
GRB 010921 &0.451 & \nodata &8.24 &\nodata   &0.00 &-19.4 &0.70\tablenotemark{g} &9.56$^{+0.09}_{-0.11}$ \\ 
GRB 091127\tablenotemark{h} &0.490 &\nodata &$\sim$8.40 &8.20  &0.04 &-18.4 &0.4 &8.3 $\pm$ 0.1 \\
GRB 070612A &0.671 &\nodata &8.29 &\nodata &0.64 &\nodata &81\tablenotemark{g} &\nodata \\ 
GRB 020405 &0.691 &\nodata &8.33/8.59 &\nodata  &0.00    &\nodata  &1.61/2.05\tablenotemark{g} &\nodata \\ 
GRB 991208 &0.706 &\nodata &8.02 &\nodata &0.58   &-18.5 &3.47\tablenotemark{g} &8.85 $\pm$ 0.17 \\ 
GRB 030528 &0.782 &\nodata &$\sim$8.40 &\nodata  &$>$0.46\tablenotemark{i}  &-20.53 &$>$12.1\tablenotemark{g} &9.11$^{+0.23}_{-0.26}$ \\ 
GRB 051022 &0.807 &\nodata &8.62          &8.37 &0.50   &-21.8  &271\tablenotemark{g} &10.42 $\pm$ 0.05 \\ 
GRB 050824 &0.828 &\nodata &$\sim$8.40 &\nodata &$<$0.16\tablenotemark{j} &\nodata &$<$0.941\tablenotemark{g} &\nodata \\  
GRB 980703 &0.966 &\nodata &8.31/8.65 &\nodata  &0.00 &-21.4 &9.9/13.6\tablenotemark{g} &9.83 $\pm$ 0.13 \\ 
\enddata	     
\tablenotetext{a}{Metallicities have a systematic error of $\pm$0.1 dex due to uncertainties in the strong line diagnostics (Kewley \& Ellison 2008).} 	
\tablenotetext{b}{Total color excess in the direction of the galaxy, used to correct for the effects of both Galactic and intrinsic extinction.}
\tablenotetext{c} {$M_B$ values come from the literature as follows: Hammer et al.\ 2006 (GRB 980425), Christensen et al.\ 2004 (GRB 980703, GRB 990712, GRB 991208, GRB 010921), Soderberg et al.\ 2004 (GRB 020903), Rau et al.\ 2005 (GRB 030528), Margutti et al.\ 2007 (GRB 031203), Gorosabel et al.\ 2005 (GRB 030329), Castro-Tirado et al.\ 2007 (GRB 051022), Mirabal et al.\ 2007 (GRB 050826), and Wiersema et al.\ 2007 (GRB 060218).}
\tablenotetext{d} {Values for the GRB 100316D and GRB 120422A hosts are from observations of the GRN/SNe sites; for more discussion see Section 4 and Levesque et al.\ (2011, 2012a) respectively.} 
\tablenotetext{e} {Since the host of GRB 031203 is not classified as a purely star-forming galaxy, all ISM properties should be taken as approximate, given the potential unknown contribution of AGN activity.}
\tablenotetext{f} {GRB 020819 values are from observations of the host nucleus; for more discussion see Section 4.}
\tablenotetext{g} {SFR determined from the [OII] line flux and the metallicity-dependent relation from Kewley et al.\ (2004).}
\tablenotetext{h} {Values from Vergani et al.\ (2011).}
\tablenotetext{i} {Lower limit from Rau et al.\ (2005).}
\tablenotetext{j} {Upper limit from Sollerman et al.\ (2007).}
\end{deluxetable}
\clearpage

\begin{figure}
\epsscale{0.8}
\plotone{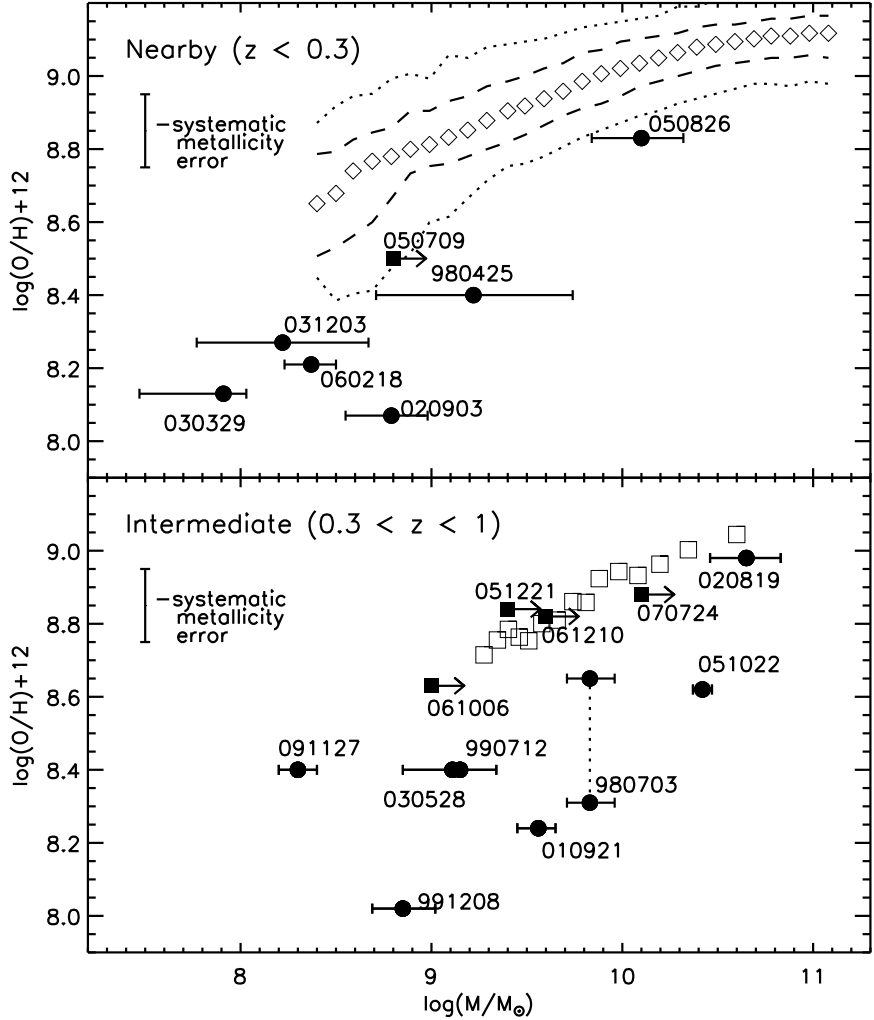}
\caption{Adapted from Levesque et al. (2010b); mass-metallicity relation for nearby ($z < 0.3$, top) and intermediate-redshift ($0.3 < z < 1$, bottom) host galaxies of LGRBs (filled circles) and SGRBs (filled squares). The GRB host galaxy data are compared to binned data for $\sim$53,000 star-forming galaxies from Tremonti et al.\ (2004; top) and 1,330 galaxies from the DEEP2 survey (Zahid et al.\ 2011; bottom). LGRB host properties are taken from Levesque et al.\ (2010a,b) and Vergani et al.\ (2011). For the SGRB hosts, metallicites are taken from Berger (2009), and stellar masses are from Leibler \& Berger (2010). All metallicities are determined using the $R_{23}$ diagnostic (Kewley \& Ellison 2008). The LGRB host sample follows its own statistically robust mass-metallicity relation (PearsonÕs $r=0.80$, $p=0.001$), while the SGRBs more cleanly trace the general star-forming galaxy population. The LGRB hosts also do not adhere to any clear metallicity cut-off despite falling below the general mass-metallicity relations for star-forming galaxies at both nearby and intermediate redshifts.}
\end{figure}

\begin{figure}
\epsscale{1.0}
\plotone{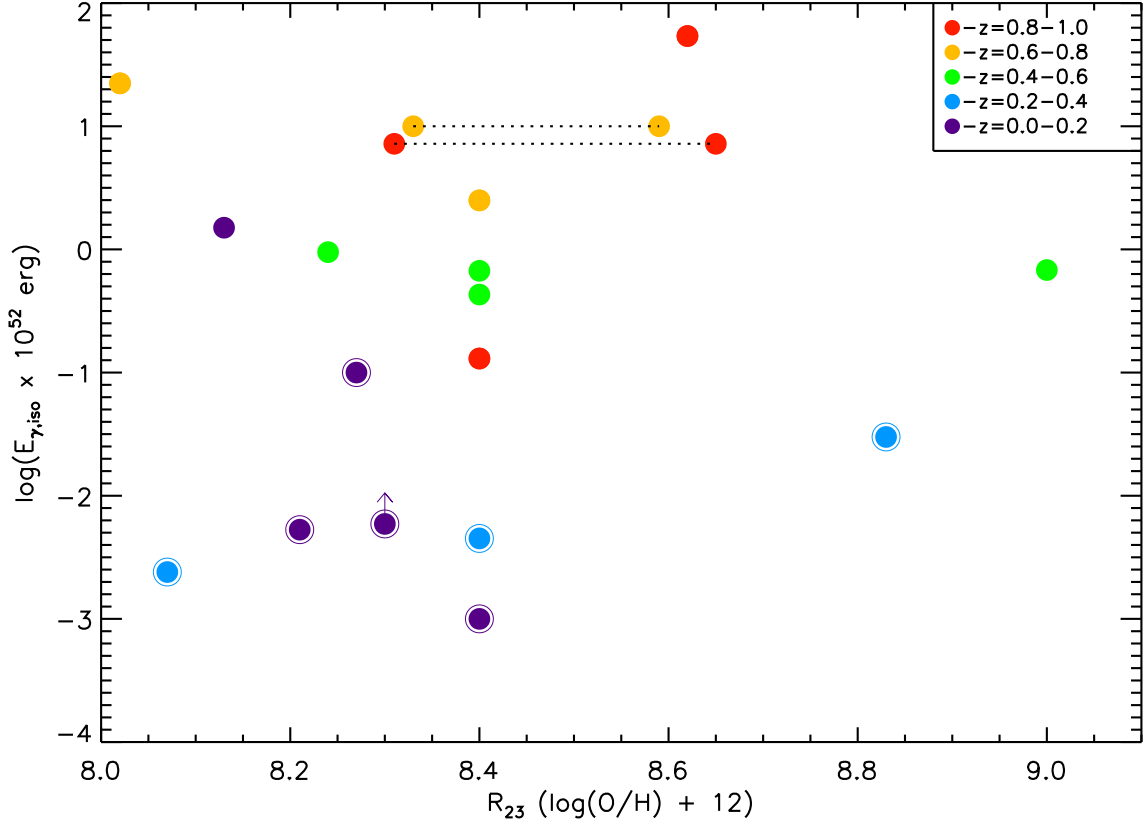}
\caption{Adapted from Levesque et al. (2010c); metallicity vs. $E_{\gamma,iso}$ for this review's sample of LGRB galaxies. The hosts have been separated into redshift bins by color in order to better illustrate redshift effects in these comparisons. All metallicities are determined using the $R_{23}$ diagnostic; for galaxies where a choice cannot be made between lower and upper branch metallicities of this diagnostic, both are plotted and connected by dotted lines. Lower limits on $E_{\gamma,iso}$ are indicated by arrows. Sub-luminous LGRBs are marked with additional outer circles. From these data Levesque et al.\ (2010c) found no statistically significant correlation (Pearson's $r = 0.08$, $p = 0.78$ assuming lower-branch metallicities, Pearson's $r = 0.10$, $p = 0.72$ assuming upper branch metallicities) between metallicity and $E_{\gamma,iso}$, a result that is at odds with past predictions and observations (e.g. Stanek et al.\ 2006).}
\end{figure}

\begin{figure}
\epsscale{0.45}
\hspace{15pt}
\plotone{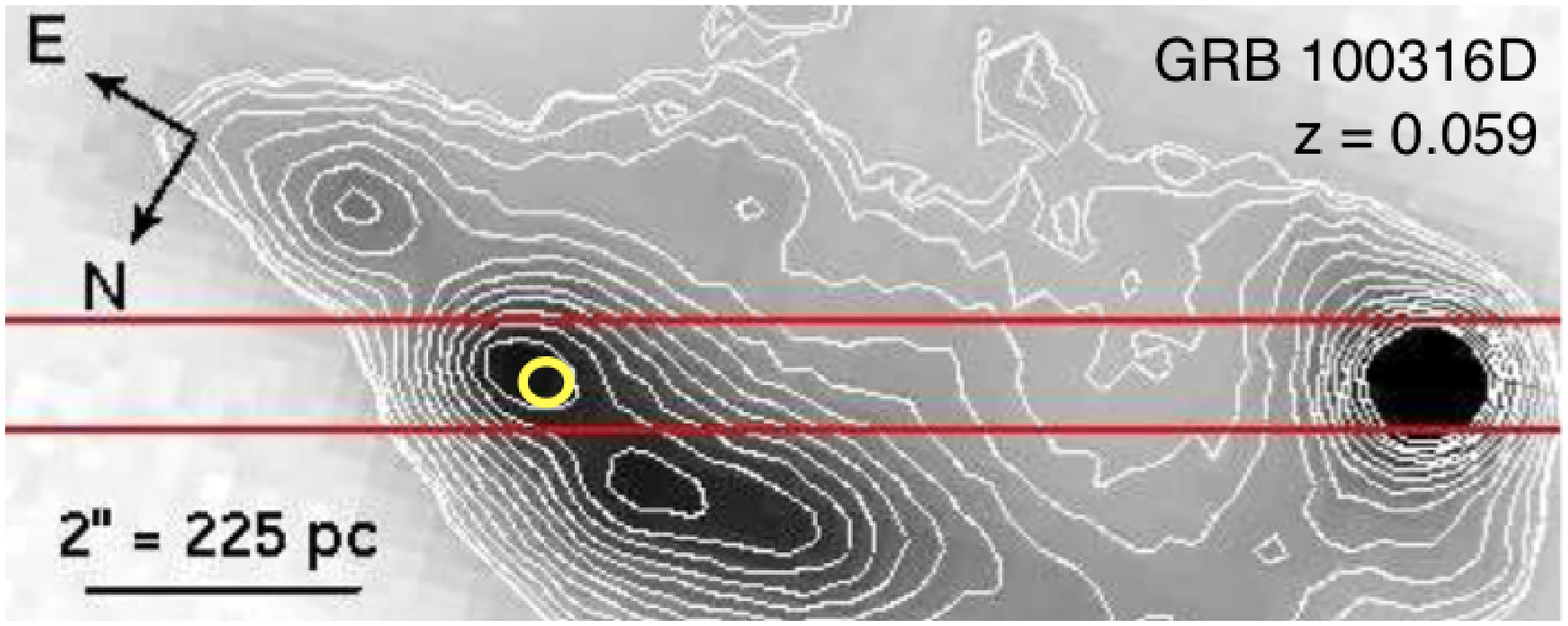}
\epsscale{0.55}
\vspace{-8pt}
\hspace{28pt}
\plotone{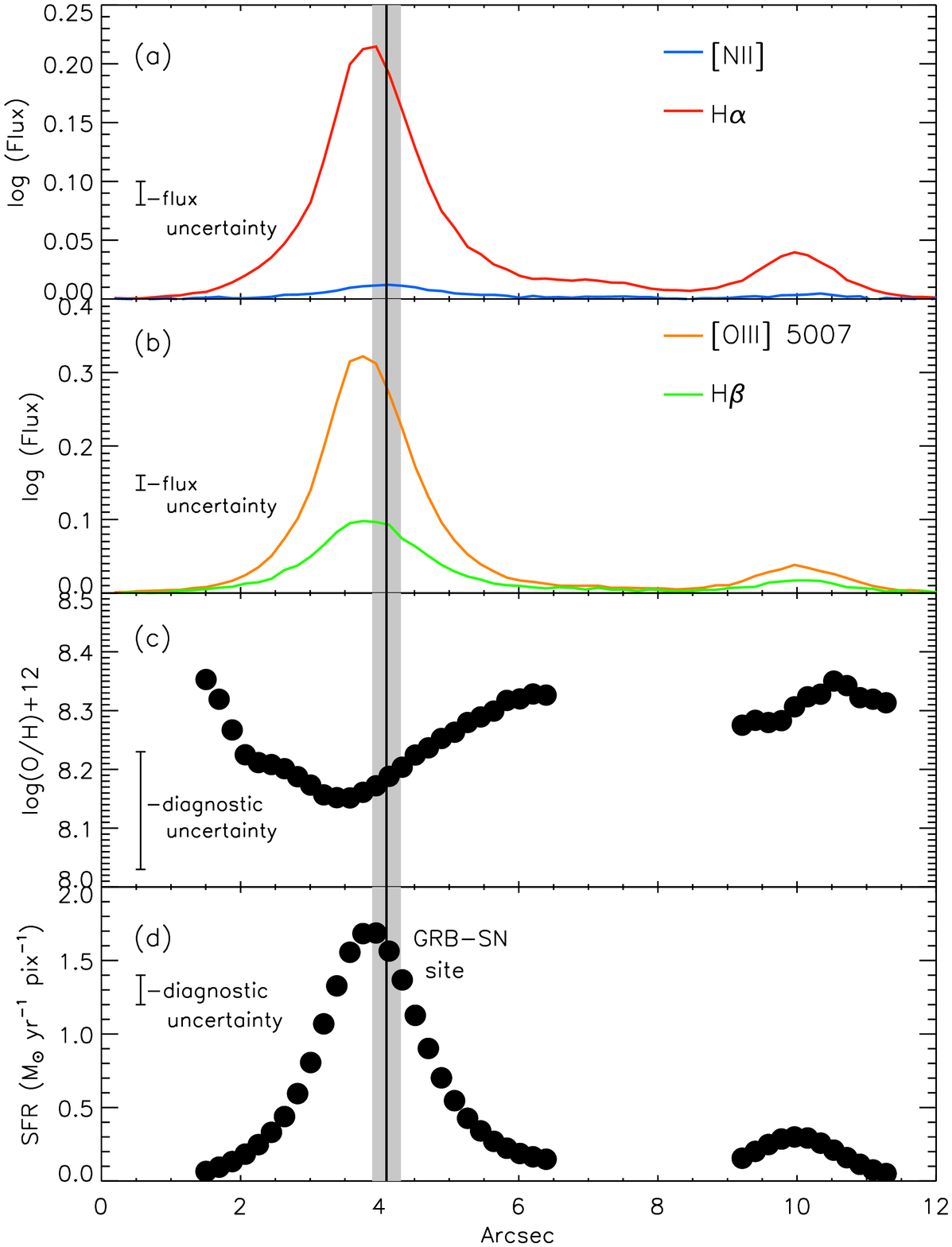}
\caption{Adapted from Levesque et al.\ (2011); photometry and spatially-resolved spectroscopy of the GRB 100316D host complex taken with LDSS3 on Magellan. Top: g-band acquisition images of the GRB 100316D host complex. Red lines illustrate the position of a 1Ó slit for acquiring spatially-resolved spectroscopy of the GRB explosion site and extended emission. The explosion site is shown as a yellow circle. Contours indicating g-band brightness. Bottom: emission line fluxes (a and b), metallicity (\ c), and SFR (d) profiles from the explosion site spectrum. Metallicity was determined based on the Pettini \& Pagel (2004) {\it O3N2} diagnostic, while SFR was determined based on the Kennicutt (1998) H$\alpha$ diagnostic. The explosion site position is indicated by the black vertical line, with errors illustrated by the gray region. In panels (a) and (b) flux is in units of 10$^{-13}$ ergs cm$^{-2}$ s$^{-1}$. From the figure it is clear that the GRB 100316D host galaxy has a relatively flat (within the errors) abundance gradient, and that the LGRB occurred near (but not precisely at) an area with a relatively high SFR and low metallicity.}
\end{figure}

\begin{figure}
\epsscale{1}
\plotone{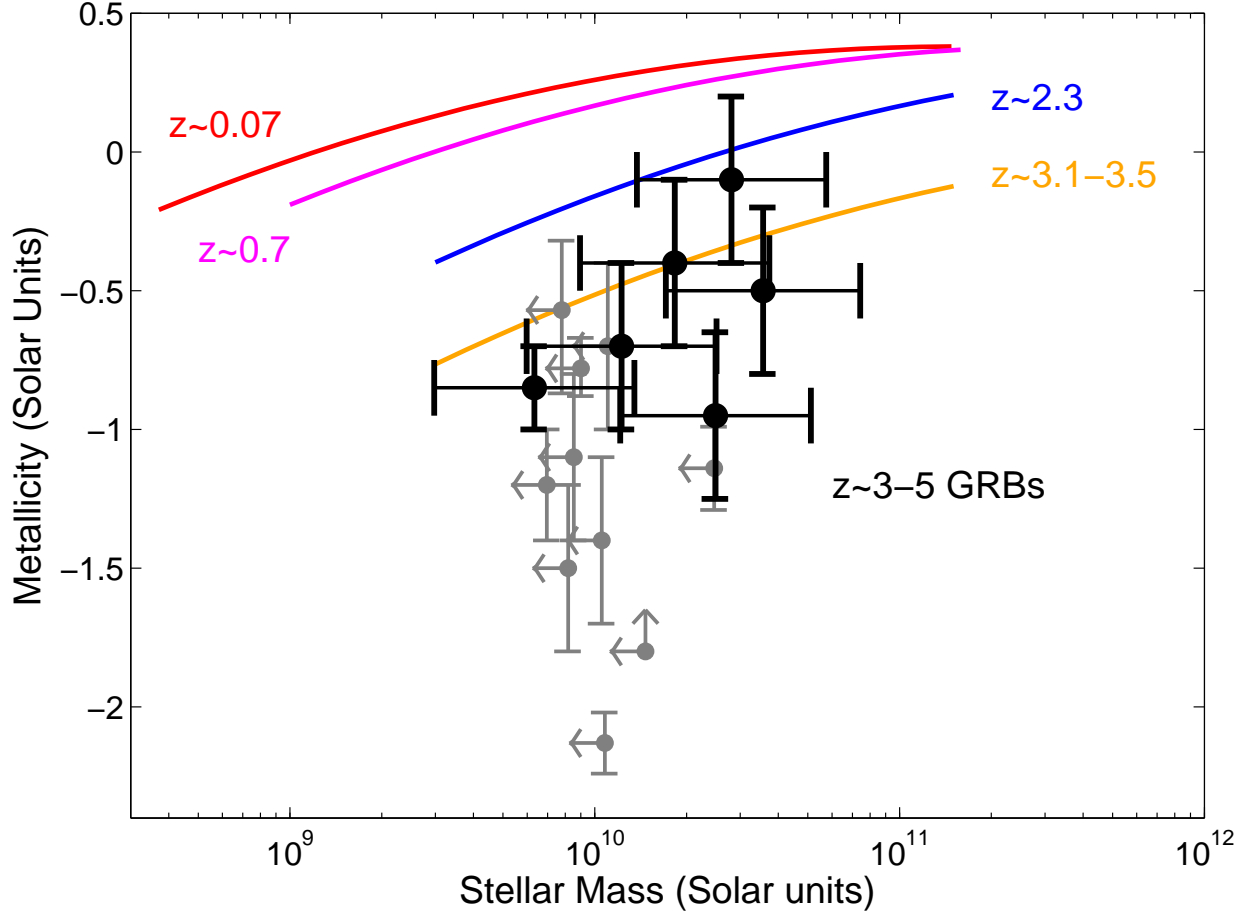}
\caption{Adapted from Laskar et al.\ (2011); stellar mass plotted as a function of ISM metallicity for their sample of  $3<z<5$ LGRB host galaxies. Black points represent metallicities from detections, while gray points are estimates determined from the correlation between Si II $\lambda$1526.7 equivalent width and metallicity found for QSO-DLAs (Prochaska et al.\ 2008). The data are consistent with a decline in metallicity at lower stellar masses, continuing the $M-Z$ relation trend seen at lower metallicities (Levesque et al.\ 2010b; Figure 1). For comparison the figure includes the $M-Z$ relations for star-forming galaxies at $z\sim0.07$ (red; Kewley \& Ellison 2008), $z\sim0.7$ (purple; Savaglio et al.\ 2005), $z\sim2.3$ (blue; Erb et al.\ 2006), and $z\sim3.1-3.5$ (yellow and filled squares; Maiolino et al.\ 2008, Mannucci et al.\ 2009). The relations at $z\lesssim2.3$ are the re-calibrated values from Maiolino et al.\ (2008). The LGRB host sample falls below all of these relations.} 
\end{figure}

\end{document}